# Application of Multi-layer Graphs In the Design of MPLS Networks

Haidara Abdalla, Dmitry Ageyev

*Abstract* – **It is suggested to use multi-layer graphs as a mathematical model in the design of MPLS networks. The application of this model makes it possible to design multi-service telecommunication systems simultaneously at several levels and to reduce the problem to the search of the minimum weight graph.**

## I. INTRODUCTION

The development of telecommunications is moving towards the usage of multi-service telecommunication systems built in accordance with the NGN concept, which are the most perspective at the moment. At the same time, one can observe that the development of technical means moves faster than the development of design methods, which is one of the most actual problems of telecommunications.

Modern telecommunication systems are multi-layer by their structure. Here, one can distinguish two types of structures: organizational and technological. Organizational structure is the structure where one can distinguish territorially distributed fragments of network performing different functions. The levels of technological hierarchy are overlay networks that use different technologies. Each logical connection on the upper level of the hierarchy uses one or several ways on a lower level.

Processes taking place on different levels of the hierarchy are closely interconnected and influence each other. The structure of network on one level has a strong influence to the characteristics of the other one. That is why when one solves the problem of structural and parametric synthesis one should regard the system being synthesed as one integral object. This multi-level and multi-aspect structure gets even more complicated when it is necessary to solve the problems of deign, when together with the complicated structure of the system being designed it is additionally necessary to take into account the different variants of the telecommunication technologies in use, their compatibility and interaction. Taking into the account of the modern multi-level systems requires the development of new mathematical models, which would allow to adequately describe the existing physical and logical connections between the elements of the system on its different levels, different types of hierarchies, and to effectively solve the problems of design.

The paper formulates the problem of synthesis of structure of MPLS network layered with the transport SDH network or WDM and suggests the method of its solving. The solving is based on the application of mathematical model of multi-layer graph.

## II. PROBLEM FORMULATION

The MPLS network considered in the thesis is multi-level and has two levels: transport network and MPLS network.

In the MPLS network, the links between the nodes are provided by the transport network. SDH networks or WDM networks can be used as a transport network depending on the required bandwidth. In such a network structure, some or all the nodes of the network support not only the technologies of the transport network, but also the MPLS technology. The node supporting the MPLS technology is called Label Switched Router (LSR). The connections between the LSR nodes are provided by physical links (or fibers depending on the used technologies of transport level) and can transit through several nodes of the transport network not supporting MPLS functionality. SLR nodes and links between them form an MPLS network layered upon the transport network.

Thus, during the planning of such networks it is necessary to determine the topology of both the networks: that of transport one and that of MPLS one. This means that one needs to determine:
- what nodes of the transport network should support the MPLS functionality;
- in what way the LSR nodes should be connected via the transport network;
- what should be the bandwidth between LSR links.

Generally, this problem is solved by the installation of MPLS equipment in some of the transport network nodes, which would provide the compromise between the cost to equip the network nodes with MPLS functions, and the costs related to the under-usage of the bandwidth of the links.

The paper [1] suggests the methods of description of modern telecommunication systems whose multilayer structure is formed by the overlay networks. The properties of multilayer networks are better taken into account if one describes the telecommunication systems by the multi-level graphs. This model allowed for the first time to show

Haidara Abdalla - Kharkiv National University of Radioelectroniks, Lenina Av., 14, Kharkov, 61166, UKRAINE,
E-mail: abdalla808@hotmail.com
Dmitry Ageyev - Kharkiv National University of Radioelectroniks, Lenina Av., 14, Kharkov, 61166, UKRAINE,
E-mail: dm@ageyev.in.ua

logical and physical connections observed in real life telecommunication systems, which allowed to take these connections into account during the planning of the systems in general, as a single integral object.

Classical approaches to the design do not take into consideration the effect of traffic aggregation on the additional logical level, and do not take into account the multi-level nature of the modern telecommunication systems. In order to solve the problem above it is necessary to use the mathematical models for multilayer networks.

## III. PROBLEM SOLVING METHOD

According to the general method for solving the problem of synthesis of multiservice telecommunication systems with the usage of multi-layer graphs, we have to synthesize the initial redundant $MLG = (\Gamma, V, E)$ In order to achieve this, we need to take the following steps:
- determine the separate layers in the synthesize network;
- represent each layer as a graph describing the interconnections on each layer;
- determine the edges forming the links between the layers;
- assign a set of values to the edges and vertices of $MLG$, characteristic for the values of the parameters of the corresponding elements and connections of the system being modeled.

Analyzing the problem formulation, one can determine the following layers:
- the bottom layer of the multi-layer graph will be the one describing the topology of the transport network;;
- the second layer on top of the bottom one corresponds to the level of MPLS network;
- the upper levels correspond to the multicast flows transferred through the network being modelled.

Synthesized multilayer MLG graph is used to solve the MPLS network design problem where the following needs to be determined:
- coordinates of LSR nodes installation;
- multicast flows transfer routes;
- transport network topology;
- links bandwidth;
- performance of the equipment installed in the nodes.

The solving of the task above can be reduced to the finding of the multi-layer minimum weight subgraph $MLG' \subset MLG$ that provides the transfer of information flows $\lambda_i \in \Lambda$ with the consideration of the requirements to the structure of the multilayer graph [2] and the flows on its edges [3] observed at the applying of constraints to the bandwidth of the edges of the multi-layer graph.

## IV. CONCLUSIONS

The paper formulates and solves the problem of MPLS network design according the minimum cost criteria. In order to solve the formulated problem it is advised to use the multilayer graph. The usage of this model allows to take into account the multi-layer nature of modern telecommunication systems formed by overlay networks, and to determine the structure of MPLS network simultaneously on two of its levels (MPLS network level and transport network level).

The paper reduces the problem of design of multiservice telecommunication system with the transferred multicast flows to the problem of finding a multilayer minimum weight subgraph with the consideration of constraints to the graph edges bandwidth.

The suggested method of the problem solving can be used in practice during the planning of NGN networks on the level of the transport network with the transfer of the multicast flows.